# Wave Tank Experiment for Sea State Monitoring with Distributed Acoustic Sensing

**Yoshiyuki Yajima[1,*], Sakiko Mishima[1], Noriyuki Tonami[1], Tomoyuki Hino[1], Shugo Aibe[1], Junichiro Saikawa[1], & Koji Mizuguchi[1]**

1. NEC Corporation

* yoshiyuki-yajima@nec.com

## ABSTRACT

Monitoring sea states across the offshore wind farm areas is essential to keep their structures safe, efficiently operate the systems, and assess the environmental effects of wind turbines. Conventional sea state sensors like buoys limit their observable coverage; therefore, installing many sensors across the wide area is necessary to obtain sufficient sea state information. However, such a situation is not practical in terms of cost. Instead, the study proposes utilising optical fibres, which is embedded in existing power cables for telecommunications on the seabed, as sea state monitoring sensors with distributed acoustic sensing (DAS). DAS is a vibration-sensing technology along optical fibres based on the Rayleigh backscattering of the injected laser. It measures the dynamic strain of the optical fibre in real time at each spatial bin, which is called a "channel" along the fibre. In power cables on the seabed, time-varying water pressure due to waves is expected to exert dynamic strain. This hypothesis motivates us to validate whether the application of DAS for power cables can estimate sea state, such as wave period, height, and the direction of arrival. Hence, the authors carried out a wave tank experiment with a programmable wave generator. An actual power cable is installed under the same condition as the bottom-mounted offshore wind turbines. The experimental results show that (i) the wave period can be accurately estimated from the frequency-domain analysis. (ii) The strong linearity between DAS vibration power and the wave height is found. (iii) The direction of arrival of waves can be estimated with the error of 1.5° when there are at least two laying angles of the cable in parallel with the estimation of wavelength. These outcomes promote the feasibility of utilising the existing power cables across offshore wind farms as sea state monitoring sensors.

## 1. INTRODUCTION

Sea state monitoring is essential for offshore infrastructures to efficient and effective operations and maintenance (O&M). For instance, structures on the sea always suffer from external forces from waves. For offshore wind turbines, sea state affects the operation of crew transfer vessels (CTVs) because crews cannot safely move to wind turbines when sea state is not calm. Underwater, power cables are also always shaken by waves, leading to fatigue and breakage of the cable. Since the operation of offshore wind farms is suspended until the power cable is fixed, it is a fatal issue. Hence, it is necessary to monitor sea state in real time.

However, conventional sea-state monitoring sensors have issues for offshore wind farm O&M. For example, buoys, one of the most common sensors, have a limited observable area. Thus, many buoys must be installed to cover the entire area of the offshore wind farm. This limitation increases the installation and maintenance costs. Other novel sensors, such as airborne LiDAR [1,2,3] and synthetic aperture radar (SAR) satellites [4,5], are proposed for sea-state monitoring. Nevertheless, they lack the real-time property; sea state can be monitored only when the airplane or satellite is above the offshore wind farm.

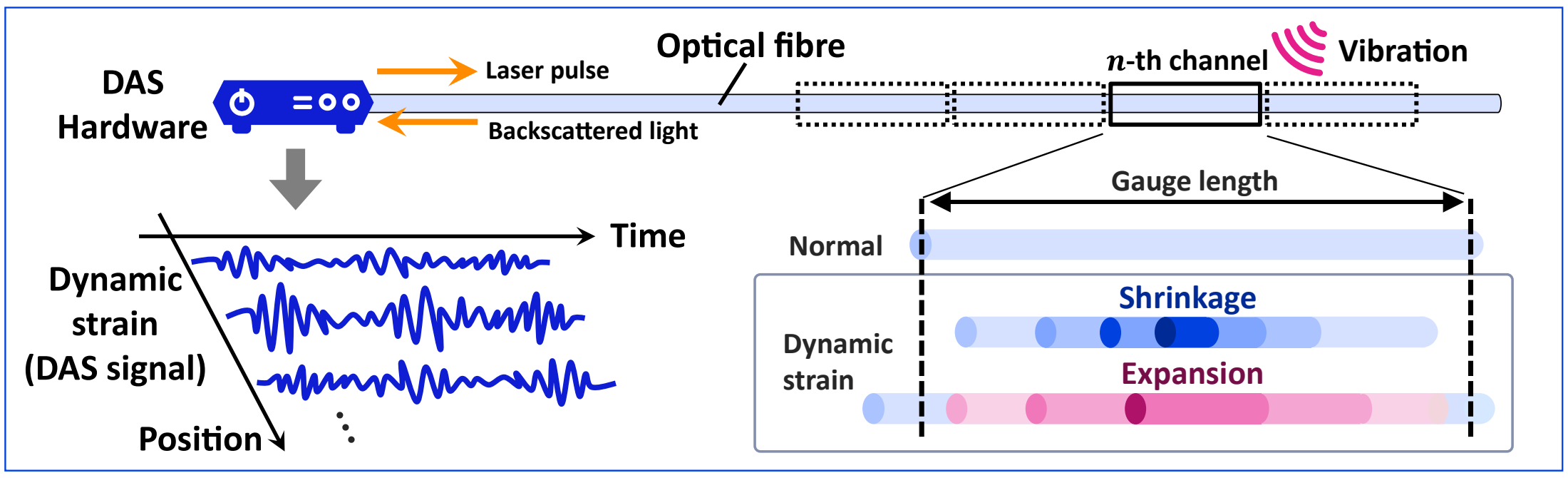

**Figure 1.** Overview of DAS.

To overcome the limitations of the existing sensors, distributed acoustic sensing (DAS) is a promising approach for a novel sea-state monitoring sensor. Figure 1 shows the schematic view of DAS. DAS measures dynamic strain in real time at each Gauge-length segment of the optical fibre, called "channel", based on Rayleigh backscattering light of incident laser [6]. In other words, DAS obtains spatiotemporally continuous vibration signals along the optical fibre. Applications for sea-state monitoring and oceanography have been reported. For example, vibration signals obtained from DAS can monitor tsunamis [7] and ocean currents [8,9] using optical fibres embedded in submarine cables. Signal processing of DAS also detects vessels under wave effects [10].

In power cables for offshore wind farms, the optical fibre for telecommunication and control is usually embedded. Recent studies suggest that DAS with the optical fibre enables health monitoring of power cables by applying multi-channel array signal processing [11]. This fact motivates us to utilise the existing power cables with DAS as a sea-state monitoring sensor instead of installing additional sensors, in addition to the application of cable monitoring. Hence, this study verifies whether sea state monitoring, in particular, wave period, height, and direction of arrival (DOA), which are fundamental quantities, can be estimated in the wave tank as a testable environment with ground truth [12].

## 2. WAVE TANK EXPERIMENT

Figure 2 (a) shows the appearance of the wave tank for the experiment. It is 50-m long, 8-m width, and 4.5-m depth with the wave generator. The controllable parameters of wave generation are height and period. The adopted wave parameters, in addition to DAS measurement parameters, are listed in Table 1. The terminal of the wave tank opposite to the wave generator sets the absorber of waves to minimise the effects of reflected waves. Waves are generated for two minutes. When the generator ran again after the previous trial, we waited until the water surface became stable.

Figure 2 (b) and (c) show the installation layout of the power cable. The cable is fixed to the aluminium frame where the wind turbine and seabed section are replicated in a vertical and the horizontal frame section, denoted as the sections "I" and "II" in Figure 2 (b). The DOA of generated waves can be controlled by changing the angle of the frame with respect to the wave tank as shown in Figure 2 (d). In the experiment, the angle is set to be –5° and –20°. The power cable is 600-V three-core CV cable armoured by galvanized iron wires as shown in Figure 2 (e). The diameter is 43 mm and the weight is 4.120 kg/m. The optical fibre cable is embedded in the power cable.

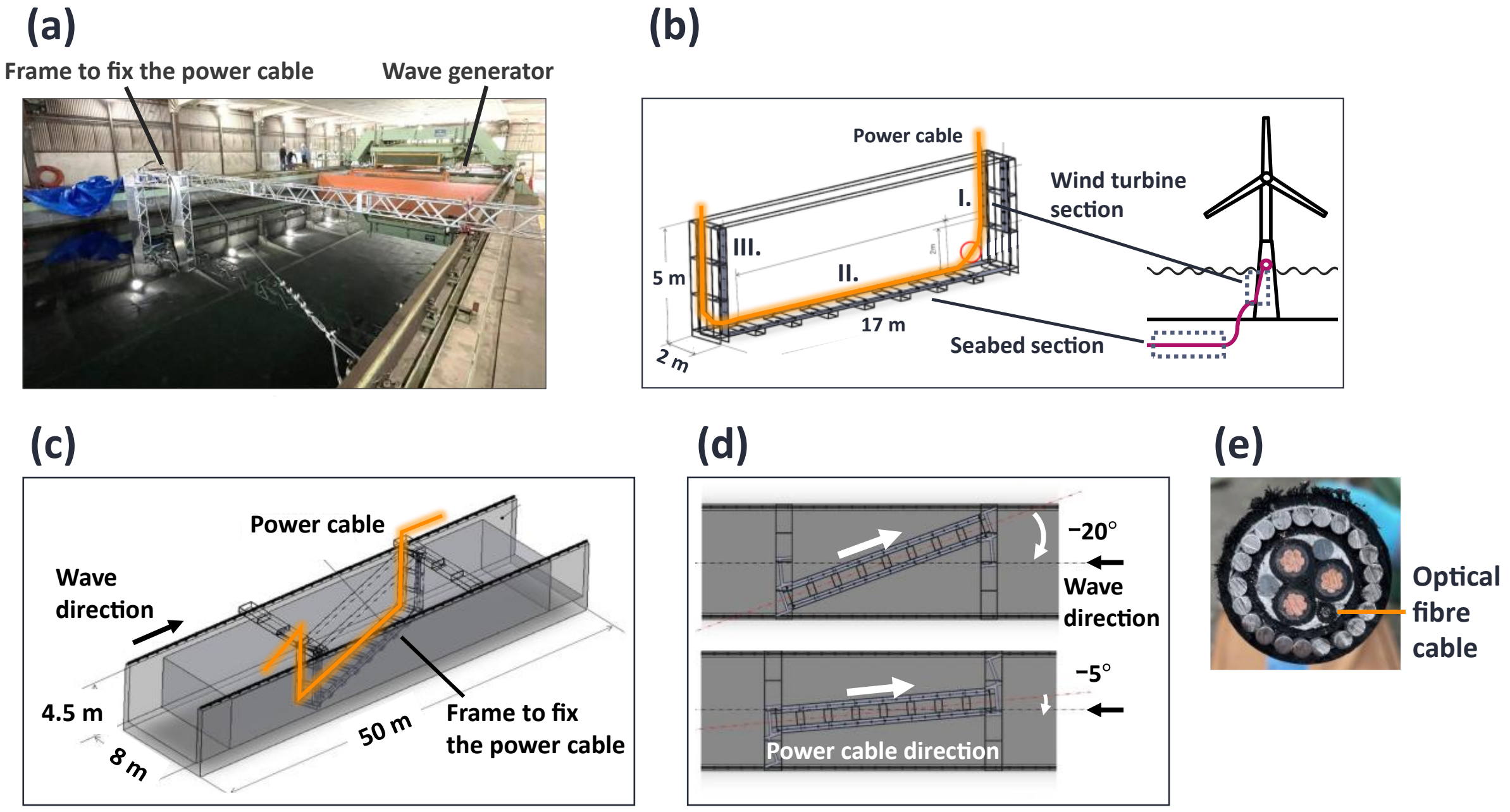

**Figure 2.** (a) Appearance of the wave tank. (b) Replicated installation conditions of the offshore wind turbine and frame dimensions. (c) Cable and frame installation conditions in the wave tank. (d) Two patterns of the wave DOA (anticlockwise to the cable direction is positive). (e) The cross-sectional view of the cable.

**Table 1.** List of generated wave parameters and DAS measurement parameters.

| Wave parameters | Value |
|---|---|
| Height [cm] | 15, 30, 40 |
| Period [s] | 1.25, 2.5 |
| Direction of arrival [deg] | –5, –20 |
| **DAS measurement parameters** | **Value** |
| Optical pulse width [m] | 2.0 |
| Gauge length [m] | 1.6 |
| Spatial sampling interval [m] | 0.80 |
| Sampling frequency [Hz] | 2000 |

# 3. RESULTS

Figure 3 shows the spatiotemporal dynamic strain observed by DAS at the beginning of 30 seconds under the condition of the wave height 30 cm, period 2.5 seconds, DOA –20°. Generated waves propagate from the top left to the bottom right in the figure. The ranges of 156–161, 161–176, and 176–181 m correspond to the vertical section of the wave-absorber section, horizontal section, and vertical section of the wave generator side, respectively. Since the vertical frames face wave fronts, its impact is directly transmitted to the frames, leading to higher dynamic strain in these sections than that in the horizontal section.

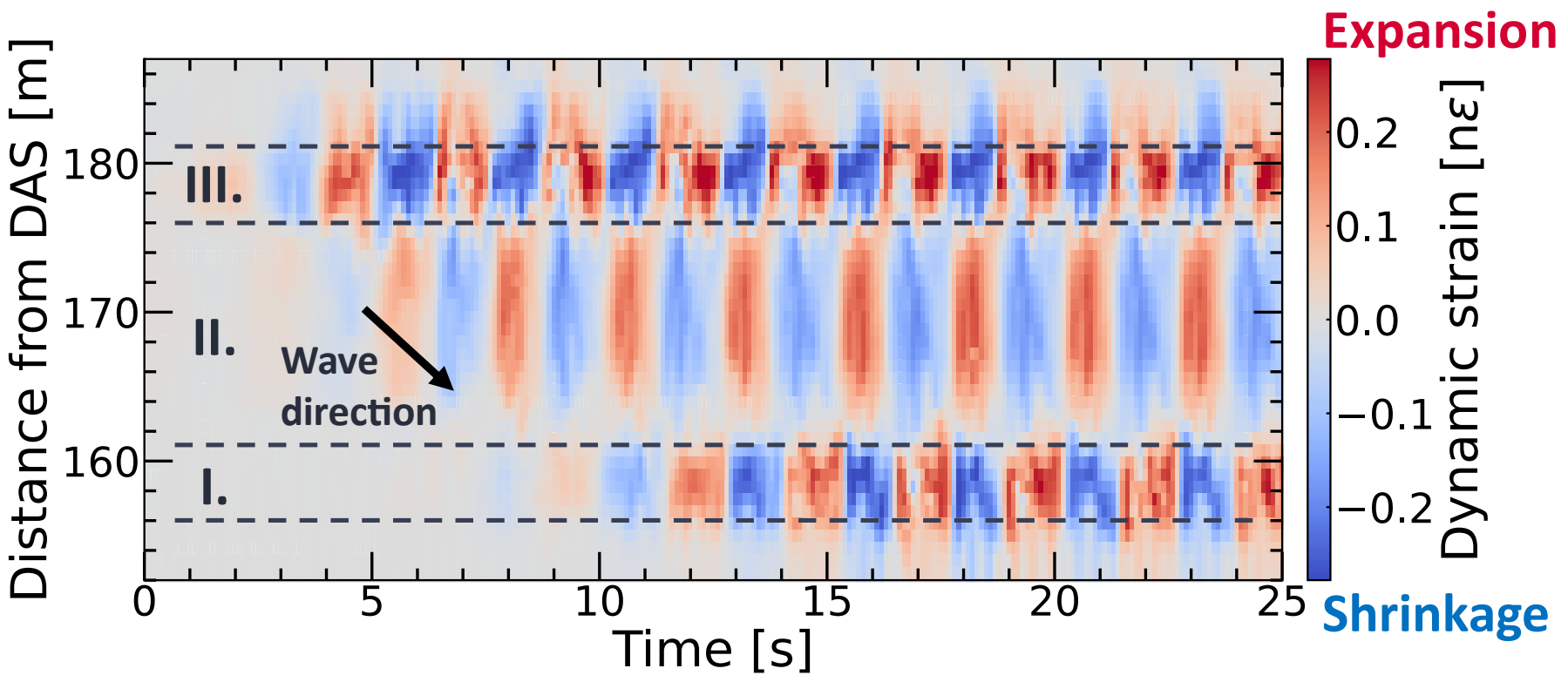

**Figure 3.** Spatiotemporal dynamic strain observed by DAS. The low-pass filter with the cutoff frequency of 3 Hz is applied. Sections "I.", "II.", and "III." correspond to those shown in Figure 2 (b).

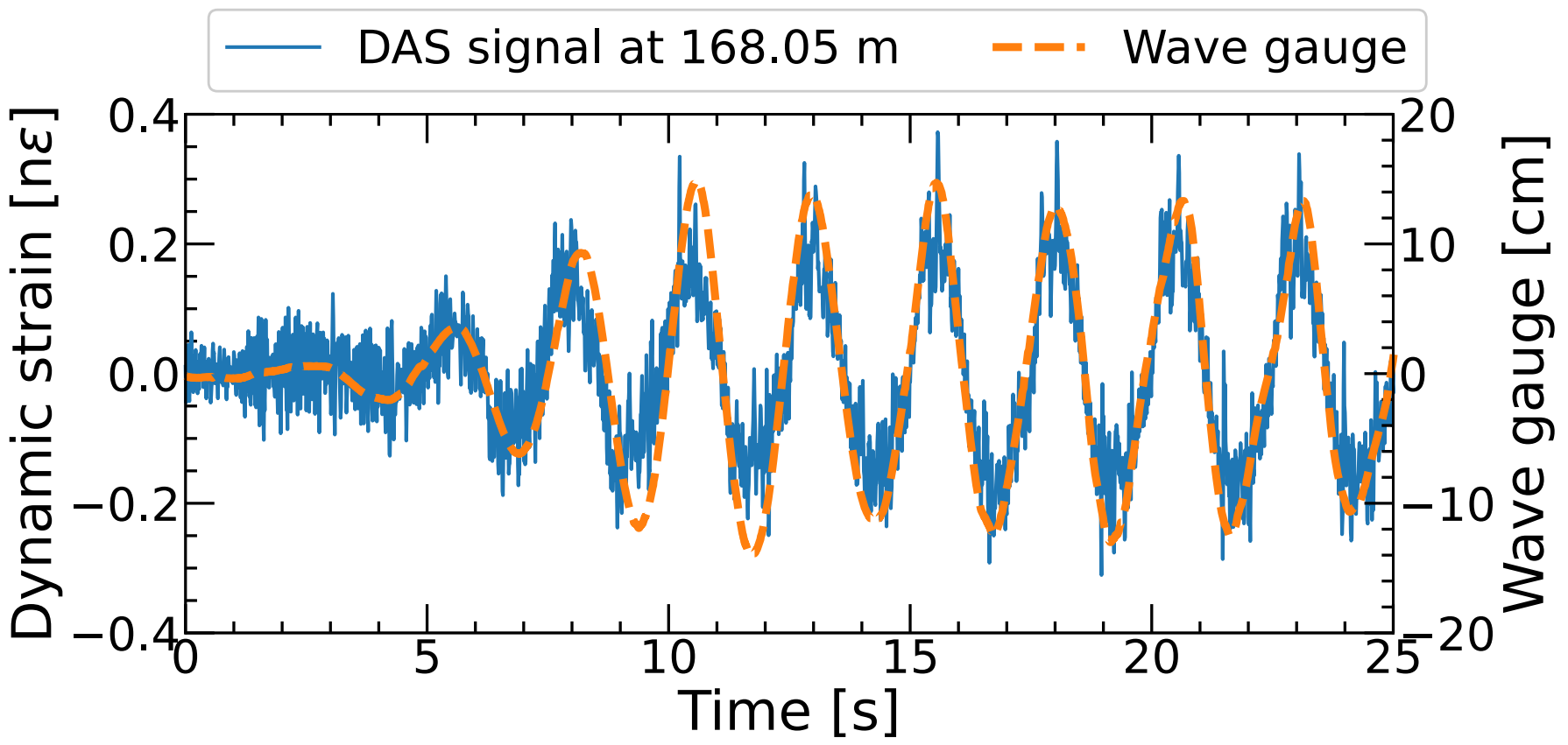

**Figure 4.** Comparison of observed dynamic strain at 168.05 m (blue solid line) and the wave gauge measurement (orange dashed line).

Figure 4 shows the comparison of dynamic strain as DAS signal at 168.05 m in the horizontal frame section and the wave gauge measurements. They have high similarity; the correlation coefficient is 0.89. Since the wave gauge measures water surface whereas the observed location of the dynamic strain is the bottom of the water, the high correlation implies that water pressure changes by waves exert influence on the power cable, resulting in highly correlated dynamic strain of the optical fibre. Hereafter, the horizontal frame section is used for the analysis, as the concept of this study assumes the use of the seabed power cable in offshore wind farms.

# 4. WAVE PARAMETER ESTIMATION

## 4.1 Wave Period

Wave period is estimated from the fast Fourier transformation (FFT) of the DAS signal at the 168.05-m channel shown in Figure 4. Figure 5 shows the power spectrum densities (PSDs) under the conditions of the wave periods 1.25 and 2.5 seconds. Wave height is fixed to be 30 cm. PSDs are derived from the Welch's method with the time window of 1 minute. The wave periods are estimated from the highest peaks at 0.403 and 0.793 Hz; other peaks are harmonics components due to vibration of the structure. The estimated results are 1.26 and 2.48 seconds, corresponding to the errors of 0.825% and 0.703% for the ground-truth wave periods of 1.25 and 2.5 seconds, respectively.

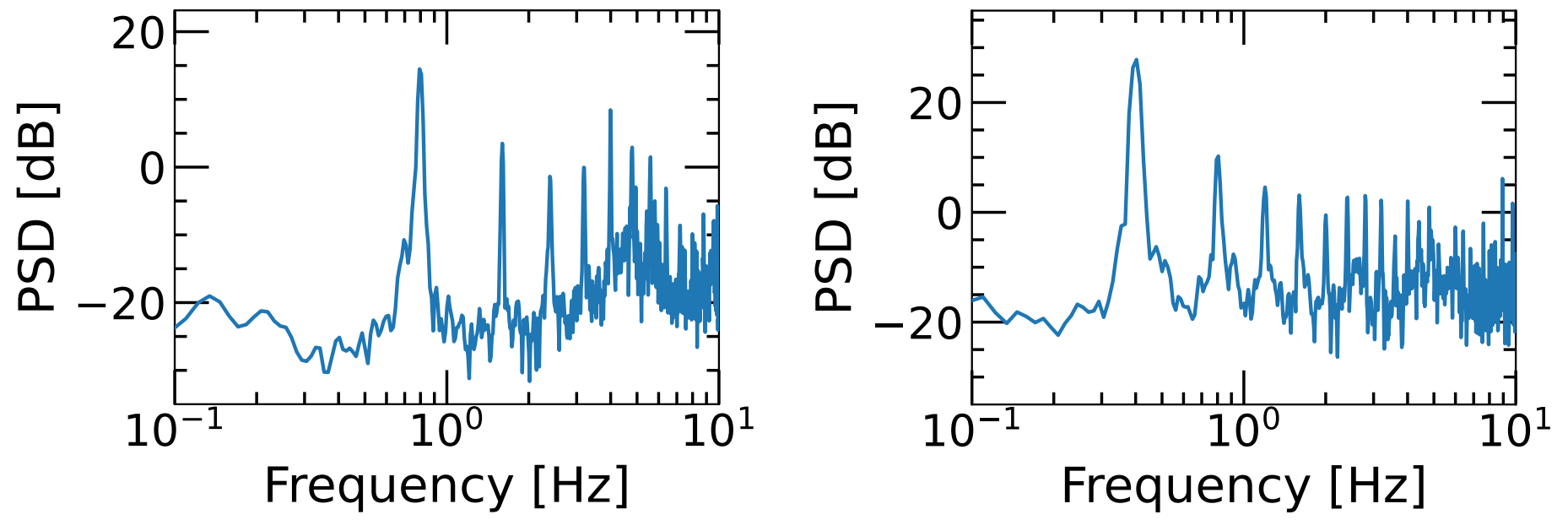

**Figure 5.** Power spectrum density of DAS signal at 168.05 m in the wave period of (left) 1.25 seconds and (right) 2.5 seconds.

## 4.2 Wave Height

This study investigates that whether wave height can be estimated from the dynamic strain observed by DAS. Therefore, the correlation between actual wave height and mean dynamic strain is examined. Figure 6 (a) shows the violin plot of root mean squares (RMSs) of dynamic strain, i.e., vibration signal power, at each wave height. The reason the statistics of RMS are investigated is that observed dynamic strain is always varied even when it is averaged over the wave period. Prior to deriving the RMS, the signal, which is 2 minutes long, is divided into 10-second intervals and the RMS is calculated for each interval. Namely, 12 samples of RMS are derived from the 2-minute signal. This process is applied to DAS channels in the range of 161–176 m, corresponding to the horizontal frame.

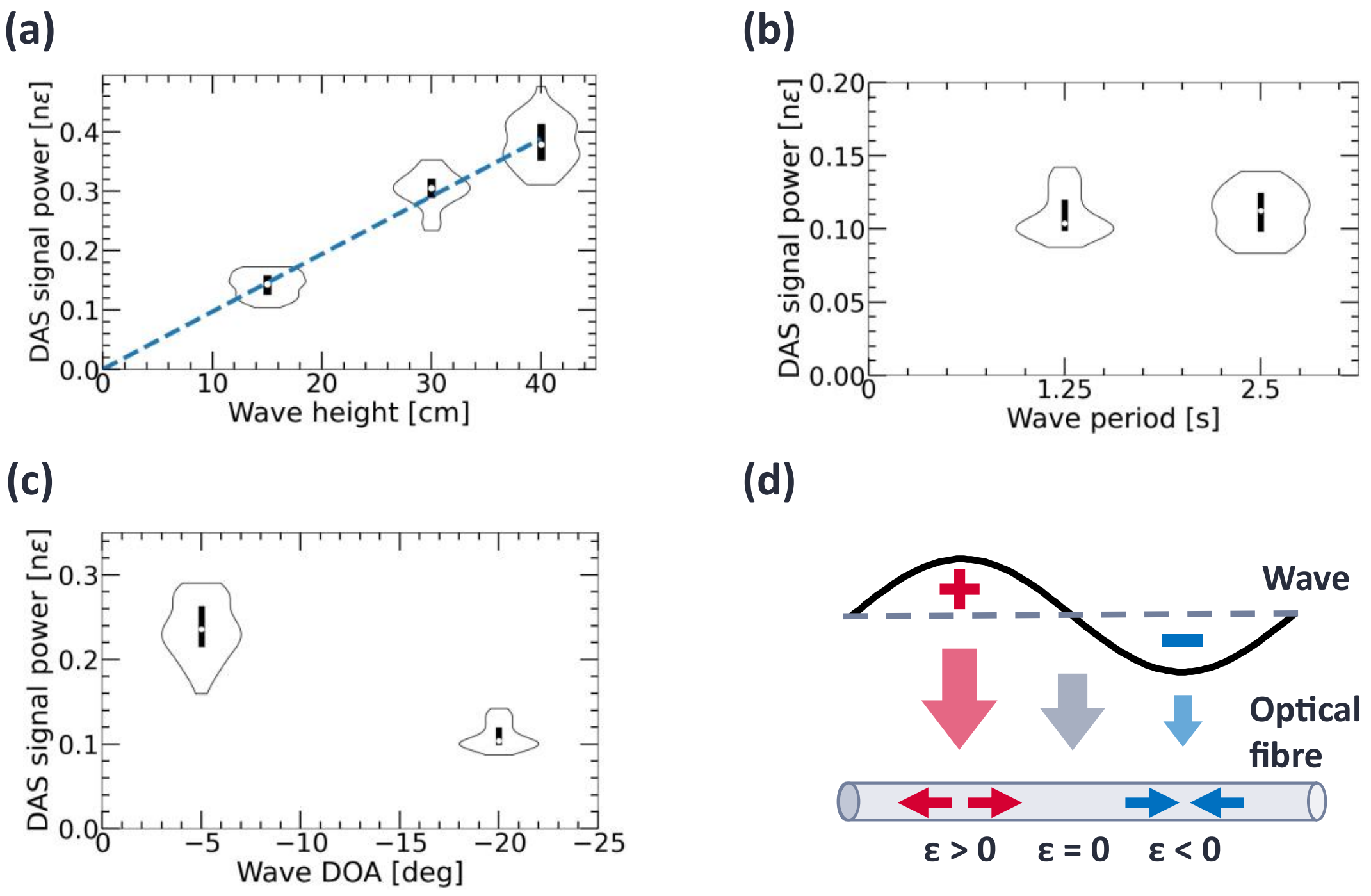

**Figure 6.** (a) Violin plot of mean dynamic strain (i.e., DAS signal power) at each wave height. The thick black lines and open circles are IQRs and medians, respectively. The blue-dashed line is the regression line of the medians. (b) The same as (a) but for the wave period. (c) The same as (c) but for the wave DOA. (d) The schematic view of the causation for dynamic strain of the optical fibre and wave crests and troughs.

Figure 6 (a) clearly shows that the RMS of dynamic strain is highly correlated with wave height. The root mean square percentage error (RMSPE) between the linear regression line of the medians derived from the least squares method with the fixed vertical intercept of zero and the medians of dynamic strain RMS is only 1.78%. This strong linearity indicates the feasibility to estimate the wave height once the calibration between observed dynamic strain and the actual wave height measured by a wave gauge is done.

The dependency of wave period and DOA on dynamic strain is also examined. Figure 6 (b) shows the violin plot of dynamic strain at each wave period with the fixed wave height. The interquartile ranges (IQRs) overlap, and the medians differ by only 9.41%. Namely, observed dynamic strain is independent of wave period. Figure 6 (c) shows the same of the panel (b) but for the DOA dependency. Even though the wave height is the same, the observed dynamic strain is higher at the DOA –5° than that of –20°. That is, DAS data has directionality.

It can be interpreted as follows. Since DAS measures the dynamic strain of the optical fibre, the influence of external forces is highest when the direction of the forces is parallel to the optical fibre as previous studies reported (e.g., [13, 14]). In contrast, forces perpendicular to the fibre have the minimum effect (note that dynamic strain is not zero in this case because of Poisson's ratio). Therefore, dynamic strain is maximum when the wave DOA is 0° and ±180° and minimum when the DOA is ±90°, assuming that the external forces by water pressure on the bottom of the water are outward beneath the wave crests and inward beneath the wave troughs, as shown in Figure 6 (d). These results support the hypothesis that the observed dynamic strain of the optical fibre is affected by only the water pressure due to water surface variation of waves, as indicated in Figure 4.

### 4.3 Wave Direction of Arrival

Since DAS can be regarded as a one-dimensional sensor array along the optical fibre, wave DOA can be estimated from beamforming, a directivity-controlling method which uses phase delays in array antennas [15]. Figure 7 (a) shows the schematic view of beamforming. The application of beamforming requires either wavelength or phase velocity; however, both are unknown for sea waves in general. Therefore, this study proposes simultaneously estimating DOA and either wavelength or phase velocity using beamforming across two fibre installation layouts. Hereafter, the cable layouts at –20° and –5° are noted as "C1" and "C2", respectively. In addition, a virtual cable layout shown in Figure 7 (b) is assumed. Although the DOA is unknown in practical, one can know that the DOA difference between C1 and C2 (C2–C1) is always 15° regardless of the DOA because it can be estimated from the cable layout map. Based on this condition, the DOA and wavelength are estimated.

Figure 7 (c) and (d) show wavelength as a function of DOA and DOA difference of C2–C1 estimated from beamforming, respectively. Here, DAS channels in the range of 161–176 m, corresponding to the horizontal frame, are applied for beamforming. Since it is already known that the DOA difference of C2–C1 is 15°, the wavelength of the waves is approximately estimated as 8.47 m according to Figure 7 (d). Based on the wavelength, the DOA for C1 and C2 are estimated as –21.4° and –6.53°, respectively, according to the dotted lines in Figure 7 (c). The resultant estimation error is 1.40° and 1.53°. In summary, this study verified that the wave DOA can be estimated using DAS and beamforming with the error of ~1.5°.

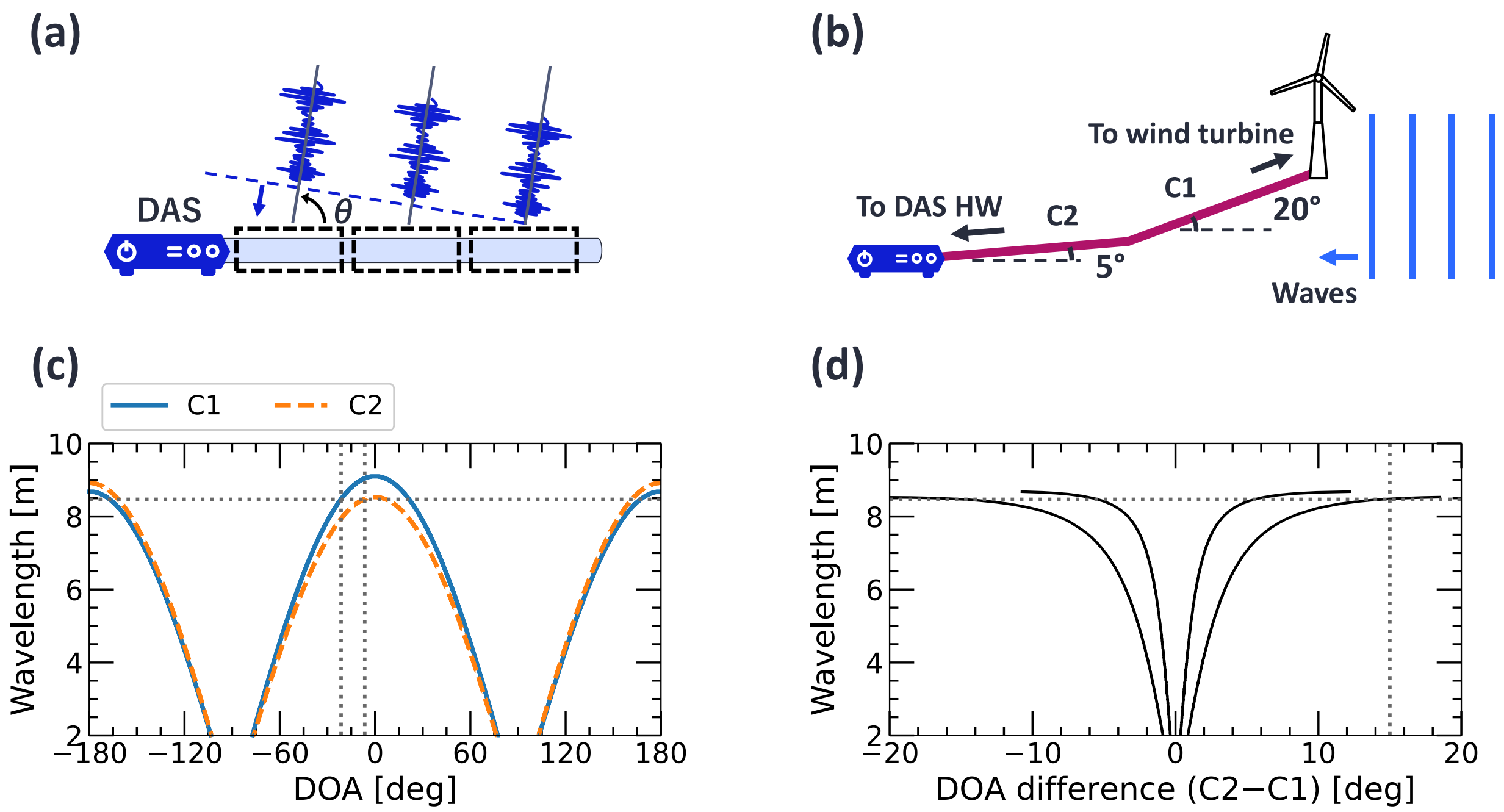

**Figure 7.** (a) Schematic view of DOA estimation with beamforming. (b) The top view of the assumed cable layout to estimate DOA. (c) Wavelength as a function of DOA estimated from beamforming for the cable layouts C1 and C2. (d) Wavelength as a function of DOA difference of C1 and C2 estimated beamforming.

## 5. SUMMARY

This study verifies the feasibility of sea-state monitoring using power cables of offshore wind farms with DAS in the wave tank experiment. DAS is a vibration-sensing technology which observes dynamic strain in each segment of the optical fibre. Namely, DAS obtains spatially continuous vibration signals as dynamic strain along the optical fibre in real time. This study demonstrates that the optical fibre embedded in the power cable is utilised as an array of wave-monitoring sensors with DAS to reduce installation and maintenance costs of conventional sensors. The power-cable installation conditions in the offshore wind farm environment, the seabed and wind turbine sections, are replicated in the experiment. The key results are listed below.

- Dynamic strain of the optical fibre in the seabed section is highly correlated with the wave gauge.
- Wave period is estimated with the error of ~0.8% with the Fourier analysis.
- The strong linearity between mean dynamic strain and wave height, with the deviation of 1.78%, is observed. Also, it is confirmed that the mean dynamic strain is independent of wave period. These facts indicate that the observed dynamic strain is affected only by wave-induced water pressure changes. It is feasible to estimate wave height using DAS once the calibration between dynamic strain and conventional sensors, such as buoys, is carried out.
- Wave direction of arrival (DOA) is estimated with the error of ~1.5° when beamforming is applied. The application of beamforming requires either wavelength or phase velocity, whilst they are usually unknown for sea waves. Therefore, this study proposes to estimate both DOA and wavelength using two cable installation layouts. This limitation is not unique for actual offshore wind farm environments where power cables are installed in several directions.

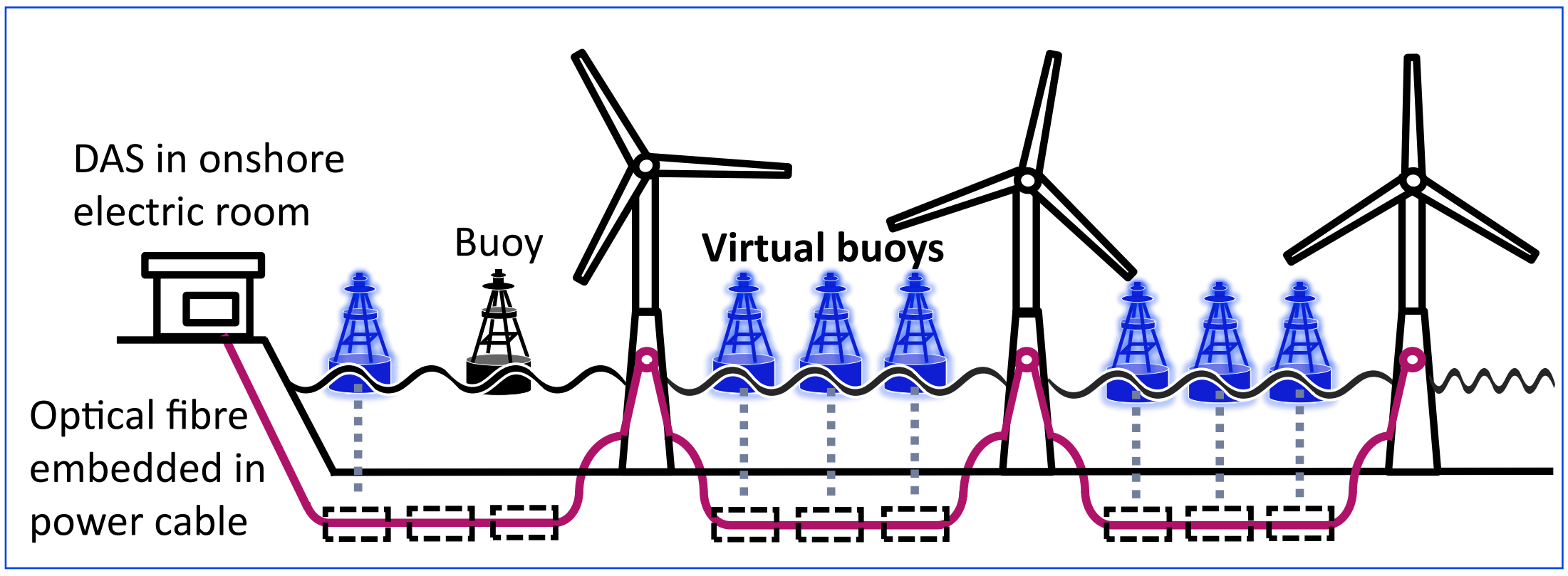

**Figure 8.** Schematic view of sea state monitoring with DAS and the optical fibre embedded in the already existing power cable of offshore wind farms.

These findings suggest that the existing power cables in offshore wind farms can be utilised as spatially continuous wave monitoring sensors like virtual buoys as shown in Figure 8. Namely, DAS enables wide-area sea state monitoring without installing additional sensors. This study demonstrates the feasibility of reducing installation and maintenance costs for sea-state monitoring sensors with DAS to cover the entire area of offshore wind farms, bringing us closer to achieving efficient O&M.

## ACKNOWKEDGEMENT

The experiment utilised the oscillation-test tank in the National Maritime Research Institute (NMRI), National Institute of Maritime, Port and Aviation Technology (MPAT) of Japan, and power cables manufactured by OCC Corporation. The authors received significant support and cooperation for the experiment from Marine Works Japan Ltd., Ocean Works Asia Inc., and NEC Networks & System Integration Corporation.